\documentclass[10pt,twocolumn,letterpaper]{article}

\usepackage[pagenumbers]{cvpr} 

\usepackage{graphicx}
\usepackage{amsmath}
\usepackage{amssymb}
\usepackage{booktabs}
\usepackage[table,xcdraw]{xcolor}
\usepackage{multirow}
\usepackage{graphicx}

\usepackage{algorithmic}
\usepackage{algorithm2e}
\usepackage[table,xcdraw]{xcolor}

\newcommand{\red}[1]{\textcolor{red}}

\usepackage[pagebackref,breaklinks,colorlinks]{hyperref}

\usepackage[capitalize]{cleveref}
\crefname{section}{Sec.}{Secs.}
\Crefname{section}{Section}{Sections}
\Crefname{table}{Table}{Tables}
\crefname{table}{Tab.}{Tabs.}
\usepackage{adjustbox}

\def\cvprPaperID{30} 
\def\confName{CVPR}
\def\confYear{2023}

\begin{document}

\title{MethaneMapper: Spectral Absorption aware Hyperspectral Transformer for Methane Detection}
\vspace{-0.25cm}
\author{Satish Kumar\\
{\tt\small satishkumar@ucsb.edu}
\and
Ivan Arevalo\\
{\tt\small ifa@ucsb.edu}
\and
ASM Iftekhar\\
{\tt\small iftekhar@ucsb.edu}
\and
B S Manjunath\\
{\tt\small manj@ucsb.edu}
\and
Department of Electrical and Computer Engineering\\
University of California Santa Barbara
}
\maketitle

\begin{abstract}
Methane (CH$_4$) is the chief contributor to global climate change. Recent Airborne Visible-Infrared Imaging Spectrometer-Next Generation (AVIRIS-NG) has been very useful in quantitative mapping of methane emissions. Existing methods for analyzing this data are sensitive to local terrain conditions, often require manual inspection from domain experts, prone to significant error and hence are not scalable. To address these challenges, we propose a novel end-to-end spectral absorption wavelength aware transformer network, MethaneMapper, to detect and quantify the emissions. MethaneMapper introduces two novel modules that help to locate the most relevant methane plume regions in the spectral domain and uses them to localize these accurately. Thorough evaluation shows that MethaneMapper achieves $0.63$ mAP in detection and reduces the model size (by $5\times$) compared to the current state of the art.  In addition, we also introduce a large-scale dataset of methane plume segmentation mask for over 1200 AVIRIS-NG flight lines from 2015-2022. It contains over 4000 methane plume sites. Our dataset will provide researchers the opportunity to develop and advance new methods for tackling this challenging green-house gas detection problem with significant broader social impact. Dataset and source code link\footnote[1]{\textit{https://github.com/UCSB-VRL/MethaneMapper-Spectral-Absorption-aware-Hyperspectral-Transformer-for-Methane-Detection}}.

\end{abstract}
\vspace{-0.20cm}
\section{Introduction}


We consider the problem of detecting and localizing methane (CH$_4$) plumes from hyperspectral imaging data. Detecting and localizing potential CH$_4$ hot spots is a necessary first step in combating global warming due to greenhouse gas emissions. 
Methane gas is estimated to contribute 20\% of global warming induced by greenhouse gasses ~\cite{kirschke2013three} with a Global Warming Potential (GWP) 86 times higher than carbon dioxide (CO$_2$) in a 20 year period ~\cite{myhre2014anthropogenic}. To put into perspective, the amount of environmental damage that CO$_2$ can do in $100$ years, CH$_4$ can do in $1.2$ years. Hence it is critical to monitor and curb the CH$_4$ emissions.
While CH$_4$ emission has many sources, of particular interest are those from oil and natural gas industries. According to the United States Environmental Protection Agency report, CH$_4$ emissions from these industries accounts to 84 million tons per year~\cite{methane-tracker-2020}. These CH$_4$ emissions emanate from specific locations, mainly from pipeline leakages, storage tank leak or leakage from oil extraction point. 

Current efforts to detect these sources mostly depend on aerial imagery. The Jet Propulsion Laboratory (JPL) has conducted thousands of aerial surveys in the last decade to collect data using an airborne sensor AVIRIS-NG~\cite{AVIRIS-NG}. Several methods have been proposed to detect potential emission sites from such imagery, for example, see~\cite{roberts2010mapping, thompson2015real, thorpe2013high, thorpe2017airborne, frankenberg2016airborne, frankenberg2005iterative}. However, these methods are in general very sensitive to background context and land-cover types, resulting in a large number of false positives that often require significant domain expert time to correct the detections. The primary reason is that these pixel-based methods are solely dependent on spectral correlations for detection. Spatial information can be very effective in reducing these false positives as CH$_4$ plumes exhibit a plume-like structure morphology. There has been recent efforts in utilizing spatial correlation using deep learning methods~\cite{kumar2020deep, jongaramrungruang2022methanet}, however, these works don't leverage spectral properties to filter out confusers. For example, methane has similar spectral properties as white-painted commercial roofs or paved surfaces such as airport asphalts~\cite{AYASSE2018386}. This paper presents a novel deep-network based solution to minimize the effects of such confusers in accurately localizing methane plumes.

Our proposed approach, referred to as the MethaneMapper (MM), adapts the DETR~\cite{carion2020end}, a transformer model that combines the spectral and spatial correlations in the imaging data to generate a map of potential methane (CH$_4$) plume candidates. 
These candidates reduce the search space for a hyperspectral decoder to detect CH$_4$ plumes and remove potential confusers.
MM is a light-weight end-to-end single-stage CH$_4$ detector and introduces two novel modules: a \textit{Spectral Feature Generator} and a \textit{Query Refiner}. The former generates spectral features from a linear filter that maximizes the CH$_4$-to-noise ratio in the presence of additive background noise, while the latter integrates these features for decoding. 

A major bottle neck for development of CH$_4$ detection methods is the limited availability of public training data. To address this, another significant contribution of this research is the introduction of a new Methane Hot Spots (MHS) dataset, largest of its kind available for computer vision researchers.
MHS is curated by systematically collecting information from different publicly available datasets (airborne sensor~\cite{gao_sensor}, Non-profits~\cite{carbon_mapper, meta_Stanford} and satellites~\cite{phiri2020sentinel}) and generating the annotations as described in Section~\ref{sec:annot}.
This curated dataset contains methane segmentation masks for over 1200 AVIRIS-NG flight lines from years 2015 to 2022. Each flight line contains anywhere from 3-4 CH$_4$ plume sites for a total of 4000 in the MHS dataset. 



Our contributions can be summarized as follows:
\begin{enumerate}
    \item We introduce a novel single-stage end-to-end approach for methane plume detection using a hyperspectral transformer. The two  modules, \textit{Spectral Feature Generator} and \textit{Query Refiner}, work together to improve upon the traditional transformer design and enable localization of potential methane hot spots in the hyperspectral images using a Spectral-Aware Linear Filter and refine the query representation for better decoding.
    \item A new \textit{Spectral Linear Filter (SLF)} improves upon traditional linear filters by strategically picking correlated pixels in spectal domain to better whiten background distribution and amplify methane signal.
    \item A new benchmark dataset, MHS, provides the largest ($\sim35\times$) publicly available dataset of annotated AVIRIS-NG flight lines from years 2015-2022.

    
\end{enumerate}

\begin{figure}[t]
\begin{center}
\includegraphics[width=\linewidth]{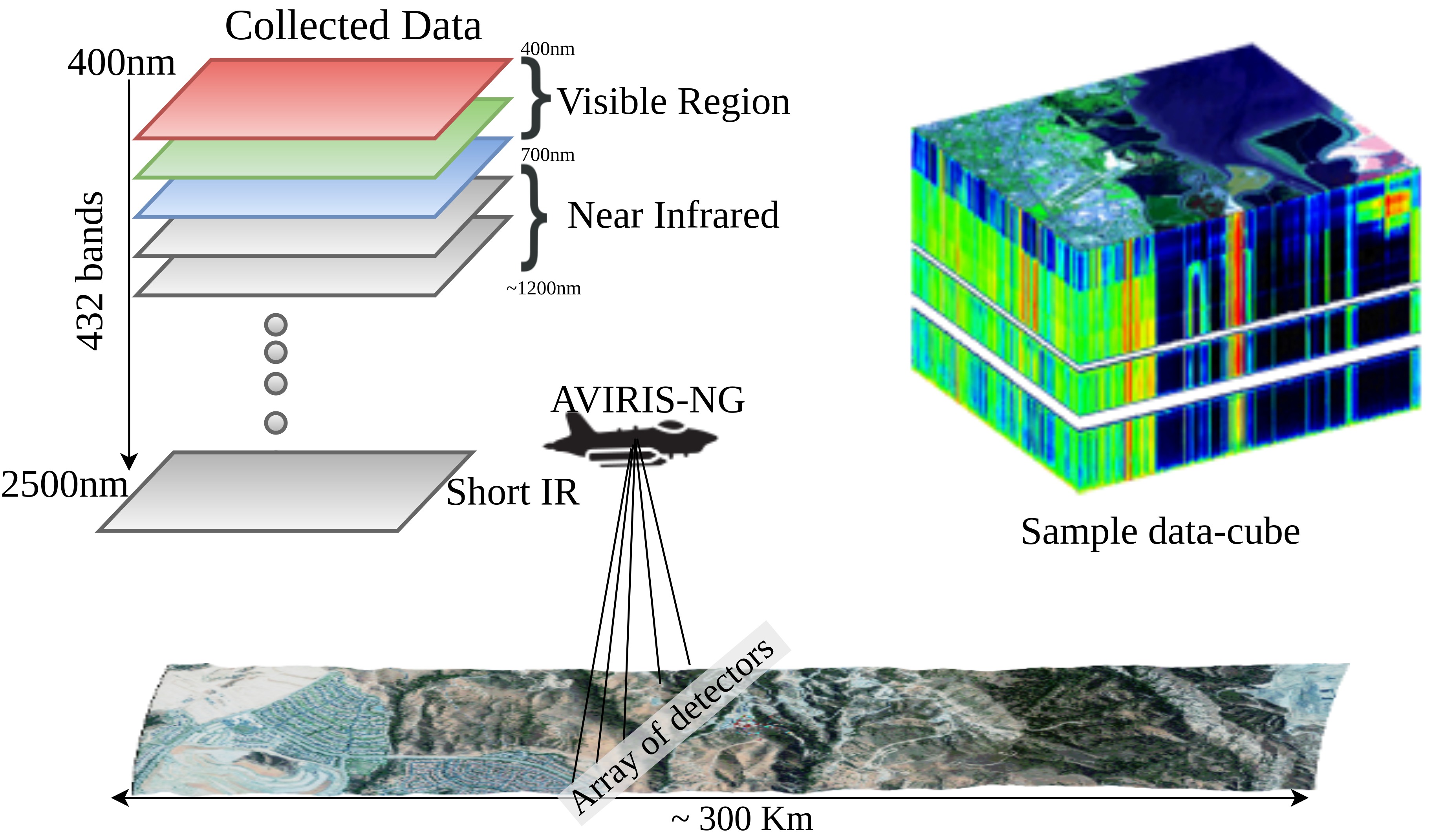}
\end{center}
\vspace{-0.45cm}
   \caption{\it Depiction of data collection process. Each flightline is $\sim300$ km long. An array of 598 sensors records data at 1.5m/pixel spatial resolution. All flightlines are ortho-corrected. Each data-cube is of dimension $\sim25000 \times \sim1500 \times 432$.}
\label{fig:data_cube}
\end{figure}

\begin{figure*}[t]
\begin{center}
\includegraphics[width=1.07\linewidth]{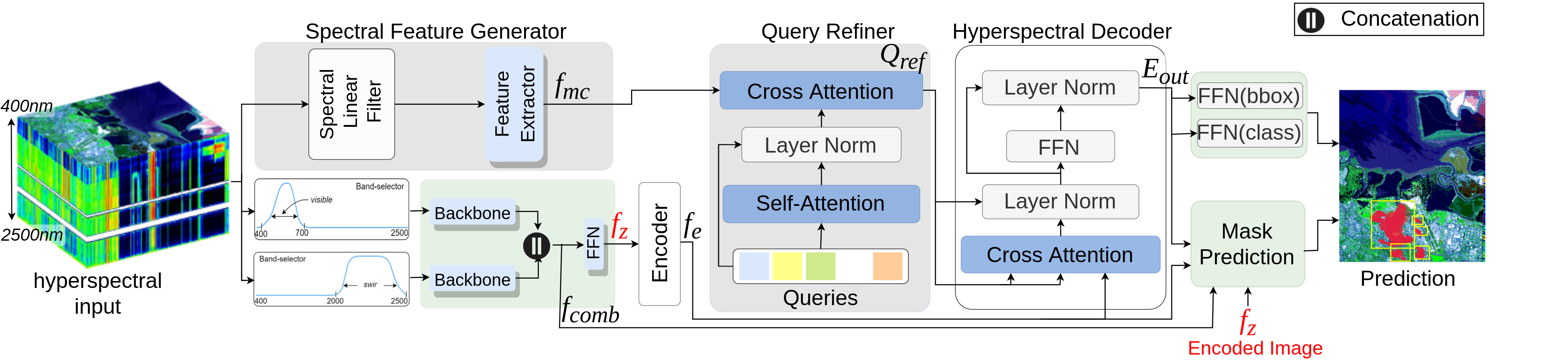}
\end{center}
\vspace{-0.2cm}
\caption{\it Overview of MethaneMapper (MM) architecture. Given a hyperspectral image, our RGB ($400nm-700nm$) and SWIR ($2000nm-2500nm$) band-pass filters passes a subset of channels in desired wavelength range and feed them to CNN backbones (ResNet) to extract features. These features are concatenated and fed to Transformer Encoder. Parallelly, our Spectral Feature Generator (SFG) modules takes in all channels of input image and generate methane candidates features. Next these candidates are sent to Query Refiner (QR) to refine queries. Then these queries decoded using encoded feature from Transformer Encoder. Finally each decoded query is used to predict a plume mask via Mask Prediction and, bounding box and class via FFNs (Feed Forward Network).
}
\label{fig:arch}
\vspace{-0.2cm}
\end{figure*} 

\section{Related Works}
Our work is at the intersection of hyperspectral data for CH$_4$ detection, deterministic linear filtering methods for spectral features and encoder-decoder based transformer. A review of the pertinent related works is given below.

There are several recent papers on detecting methane plumes from the airborne imaging spectrometer AVIRIS-NG~\cite{AVIRIS-NG}. This includes the Iterative Maximum a Posterior Differential Optical Absorption Spectroscopy algorithm (IMAP-DOAS)~\cite{frankenberg2004iterative, frankenberg2005iterative} and matched filters~\cite{roberts2010mapping, frankenberg2016airborne,thompson2015real, thorpe2013high, thorpe2017airborne}. IMAP-DOAS requires data from 2 hyperspectral sensors, one airborne and another on ground, hence not very practical for most application scenarios.  
Matched-filter based methods use background statistics to normalize the spectral signals and match with the CH$_4$ spectral signature at every spatial location (pixel-wise). This process, however, is sensitive to surface albedo and land cover with spectral absorption similar to CH$_4$, leading to spurious detections. Domain experts must then manually inspect each flight line to identify and delineate real CH$_4$ plumes~\cite{thompson2017isgeo}. To suppress the effect of false positives due to variability of elements on ground, \textit{Christopher et.al.}~\cite{funk2001clustering} and \textit{Thorbe et. al.}~\cite{thorpe2013high} introduced cluster-tuned matched filter. It involves clustering the pixels with similar spectral properties using k-means clustering. Both IMAP-DOAS and all versions of matched filters are heavily prone to false positives as the information is processed pixel-wise. 

Machine learning approaches have been used for target detection, including CH$_4$ identification, in hyperspectral imagery~\cite{gewali2018machine, hong2021spectralformer, shah2007ica, tax2004support, kumar2023situ}. 
Similar to matched-filtering, these methods do not take into account the influence of confusers on the CH$_4$ spectral signature and 
have similar issues concerning false positives. Recently introduced deep learning based H-mrcnn model~\cite{kumar2020deep} focus on capturing spatial correlation. 
H-mrcnn is an ensemble of mask-rcnn{cite} networks processing blocks of hyperspectral data. This block processing in the spectral domain is inefficient and often results in overall poor performance.
Methanet~\cite{jongaramrungruang2022methanet} is a more recent work focusing on estimating methane concentration from matched-filter data.
In this regard, our proposed MethaneMapper uses both spectral and spatial correlation to accurately delineates CH$_4$ plumes.

\noindent \textbf{Datasets:} The only dataset publicly available with annotation for CH$_4$ plume detection is JPL-CH$_4$-detection2017-V1.0 dataset~\cite{thompson2017isgeo}. It contains only 46 AVIRIS-NG~\cite{AVIRIS-NG} flight lines in the US Four-Corners region. Deep learning architectures require a large number of annotated samples, and for this reason we introduce the new MHS dataset with over $1200$ annotated flightlines and $\sim4000$ plume sites.

\section{MethaneMapper (MM) Architecture}

\subsection{Data Overview}
AVIRIS-NG hyperspectral imaging sensors capture spectral radiance values from N$_0$ (N$_0$ = 432) channels corresponding to wavelengths ranging from $400nm - 2500nm$ as shown in Fig.~\ref{fig:data_cube}. The complete hyperspectral image is represented as \textbf{x} $\in \mathbb{R}^{H_0 \times W_0 \times N_0}$ where $H_0,W_0$ are the height \& width, respectively, and $N_0=432$ is number of channels. This hyperspectral data includes a very weak signature of CH$_4$ around 2100-2400$nm$, conflated with radiations from the surrounding land cover and background clutter. A single flight-line could be over a couple miles long (about 25K pixels in one of the dimensions), with an array of sensors recording the data at 1.5m/pixel resolution. The images are orthorectified before processing. 
\subsection{Technical Overview}
Referring to Fig.\ref{fig:arch}, MM contains the following main components: (i) 2 CNN backbones to extract a compact feature representation of the spectral regions of interest from the hyperspectral image, (ii) a \textit{Spectral Feature Generator} (\textbf{SFG}), and (iii) a \textit{Query Refiner} (\textbf{QR}) in between an encoder-decoder pair (inspired by GTNet~\cite{iftekhar2021gtnet}, SSRT~\cite{iftekhar2022look}).  
The  hyperspectral image is first processed through two separate band-pass filters to select the channels in visible ($400-700nm$) and short-wave infrared (SWIR)($2000-2500nm$) wavelength regions, and are then passed through CNN backbones. Output of these backbones are concatenated together and then encoded using a transformer encoder. 

The \textbf{SFG} (Sec.~\ref{sec:SFG}) takes in all channels of the hyperspectral image and process them through a spectral linear filter. The \textbf{SFG} exploits the spectral correlation to generate methane candidates feature maps and passes them to \textbf{QR}. The \textbf{QR} (Sec.~\ref{sec:QR}) uses these methane candidates to refine the learnable queries. Our hyperspectral decoder takes the encoded features from the encoder and refined queries from \textbf{QR} to generate the embeddings. The mask-prediction layer processes these embeddings along with the feature pyramid from the backbone layers to generate the final methane-plume segmentation prediction.

\noindent These individual blocks are discussed in more detail below.


\subsection{Bandpass filtering for the Encoder}
The HSI is processed by two parallel band-pass filters; a visible wavelength ($400-700nm$) (RGB) and a short-wave infrared wavelength ($2000-2500nm$) (SWIR) band-pass filter. The RGB filter results in a 3 channel output corresponding to the normal red, green, and blue wavelengths. The SWIR generates channels, approximately 5nm apart. The filtered outputs are \textbf{x$_{rgb}$} $\in \mathbb{R}^{H_0 \times W_0 \times 3}$ and \textbf{x$_{swir}$} $\in \mathbb{R}^{H_0 \times W_0 \times 100}$. Using \textbf{x$_{rgb}$} and \textbf{x$_{swir}$}, two conventional CNN backbones (e.g. ResNet-50~\cite{he2016deep, kumar2021stressnet}) generate two feature maps respectively of size $\in \mathbb{R}^{H \times W \times N}$. Here $H=\frac{H_0}{32}, \; W=\frac{W_0}{32}$ and $N=2048$ typically. We concatenate these feature maps along channel dimension and project through a $1\times1$ convolution layer to retain channel dimension of $N$. The resulting output is \textit{f$_{comb}$} $\in \mathbb{R}^{H \times W \times N}$. 

Following the standard architecture of transformer encoder from previous works~\cite{carion2020end, iftekhar2022look, iftekhar2021gtnet, ulutan2020vsgnet, kumar2022locl}, we reduce the channel dimension of \textit{f$_{comb}$} using $1\times1$ convolution to \textit{f$_{z}$} $\in \mathbb{R}^{H \times W \times d}$ and supplement position information by adding a fixed positional embedding \textit{p} $\in \mathbb{R}^{H \times W \times d}$. The encoder consists of a stack of multi-head self-attention modules and feed-forward networks (FFN). The encoded feature map is \textit{f$_{e}$} $\in \mathbb{R}^{H \times W \times d}$:
\begin{equation}
    \textit{f$_{e}$} = \text{Encoder} (\textit{f$_{z}$}, \textit{p})
    \label{eq:encoder}
\end{equation}

\subsection{Spectral Feature Generator (SFG)}
\label{sec:SFG}
In parallel, the input hyperspectral image is processed by the \textbf{SFG} module to generate methane candidates feature map \textit{f$_{mc}$}, providing the \textbf{QR} module with spatial information to help the network delineate the methane plumes. 

The \textbf{SFG} consist of a spectral linear filter (SLF) and a Feature Extractor (e.g.ResNet-50~\cite{he2016deep}). The most common linear filtering approach for detecting CH$_4$ is to take each pixel from the input hyperspectral image \textbf{\{x$_{ij}$ $|$ x$_{ij}$ $\in \mathbb{R}^{1 \times 1 \times N_0}\}$}$_{i,j=1}^{H_0, W_0}$ and project it onto a CH$_4$ spectral absorption signature vector of same size~\cite{gordon2022hitran2020}.  This is to reduce the interference from ground terrain and amplify the CH$_4$ visibility in that pixel. Accurately modeling SLF is critical given that it is designed to reduce ground terrain interference.
To model \textbf{SLF} we use the most common approach to matched filtering from information theory~\cite{turin1960introduction}. 

\paragraph{Spectral Linear Filter (SLF):} The design of SLF is dependent on the spectral absorption pattern of CH$_4$ gas~\cite{gordon2022hitran2020} and distribution of ground terrain. Since our signal of interest, CH$_4$, is very weak, traditional methods of linear filtering~\cite{thorpe2013high, thorpe2017airborne} are not effective. The conventional methods to whiten the ground terrain noise includes calculating the covariance (\textbf{Cov} $\in \mathbb{R}^{N_0 \times N_0}$) of background by selecting a set of 10-15 adjacent columns \textbf{\{x$_{i}$ $|$ x$_{i}$ $\in \mathbb{R}^{1 \times H_0 \times N_0}\}$}$_{i=1}^{W_0}$).  However, in a given flight-line, the terrain changes frequently, from water bodies to bare soil, vegetation, buildings and other urban structures. Therefore single approximation of the covariance can not provide correct estimate of CH$_4$ and a localized context-based whitening will be more effective. 
To address this problem, we took a very simple and effective approach of doing land cover classification and segmentation~\cite{NDVI-classification, normalized-difference-vegetation-index, gao1996ndwi}, and then compute covariance per class from the land cover. More details in supplementary materials. 
This improves the quality of methane candidates in presence of confusers (materials with similar spectral absorption patterns as CH$_4$) and also in cases where CH$_4$ concentration is low. The final \textbf{SLF} design with per class covariance is:
\begin{equation}
    \textbf{SLF} (\textbf{x}_{ij}) = \frac{(\textbf{x}_{ij}-\mu_k)^T\textbf{Cov}_k^{-1}\textit{t}}{\sqrt{\textit{t}^T\textbf{Cov}_k^{-1} \textit{t}}} \; \forall\; (i,j) \in \text{class}~k
    \label{eq:mf22}
\end{equation}
where $t$ represents the spectral absorption pattern~\cite{gordon2022hitran2020} of CH$_4$ gas, 
and \textbf{Cov}$_k$, $\mu_k$ are the covariance and mean of $k^{th}$ class respectively. \textbf{x}$_{ij}$ represents the pixel in input hyperspectral image at ($i,j$) index in $k^{th}$ class. 
This operation generates a 2-D spatial CH$_4$ candidates map of size $\mathbb{R}^{H_0 \times W_0}$. Next this CH$_4$ candidates map is fed to a Feature Extractor to generate CH$_4$ candidates feature map \textit{f$_{mc}$}. Details of the land cover segmentation/classification and complete SLF derivation  are in the Supplementary materials.

\begin{equation}
    f_{mc} = \text{FeatureExtractor}(\;\textbf{SLF}(\textbf{x}_{ij})\;\forall \; i, j)
\end{equation}

\subsection{Query Refiner (QR)}
\label{sec:QR}
Next the methane candidate feature map \textit{f$_{mc}$} $\in \mathbb{R}^{H \times W \times d}$ is fed to the \textbf{QR} module along with a set of $100$ learnable queries \textit{Q} $\in$ $\mathbb{R}^{100\times d}$. The \textit{f$_{mc}$} refines the learnable queries via cross-attention mechanism. This operation provides a narrow search space for the queries. The \textbf{QR} module follows a transformer decoder-like architecture inspired from~\cite{iftekhar2021gtnet, iftekhar2022look}. 
The randomly initialized queries \textit{Q} $\in \mathbb{R}^{100\times d}$ are first passed through a self-attention layer to attend to themselves. Next, these queries attend to our methane candidates feature map \textit{f$_{mc}$} from \textbf{SFG} module through a cross-attention layer. The methane candidates feature map serves as key-values pairs in our attention architecture. The output of \textbf{QR} is \textit{Q$_{ref}$}.
\begin{equation}
    Q_{ref} = \textbf{QR}(f_{mc}, Q)
\end{equation}

\subsection{Hyperspectal Decoder}
The \textit{Q$_{ref}$} is fed to the decoder module along with encoder output \textit{f$_{e}$} to generate output embeddings. Our hyperspectral decoder follows the standard architecture with a minor difference. There are no self-attention layers, just stack of multi-headed cross attention layers. The refined queries are transformed into output embeddings \textit{E$_{out}$} $\in \mathbb{R}^{100\times d}$.
\begin{equation}
    E_{out} = \text{Decoder} (f_e, p, Q_{ref})
\end{equation}

\subsection{Box and Mask Prediction}
The decoder output embeddings (\textit{E$_{out}$}) are fed to two Feed Forward Network (FFNs) and a Mask prediction layer. The outputs of the FFNs are the bounding boxes covering each CH$_4$ plume and a confidence score corresponding to each box. The mask-prediction module follows the standard segmentation head of DETR~\cite{carion2020end}. It computes multi-head attention scores of each embedding over the \textit{f$_{e}$} (Eq. \ref{eq:encoder}), generating a low-resolution heatmap for each embedding. To make the final prediction a Feature Pyramid Network~\cite{he2017mask} like structure is used. Each heatmap is designed to capture one methane plume. A simple thresholding is used to merge the heatmaps as final segmentation mask.
\begin{equation}
    mask = \text{Mask\_pred} (E_{out}, f_e, f_{comb})
\end{equation}

\subsection{Training and Inference}
We train MethaneMapper in two stages; first we train bounding box detection corresponding to each CH$_4$ plume, and second by freezing the box detection network and training only the mask prediction module. We also trained both box and mask prediction modules end-to-end and achieved similar performance. We use a similar two-stage loss strategy for training MethaneMapper as that used in DETR~\cite{carion2020end}: first stage is the bipartite matching between the predictions and the ground truths both in bounding box  and mask prediction, and then second stage is loss calculation for the matched pairs. The bipartite matching employs the Hungarian algorithm\cite{carion2020end} to find the optimal matching between the predictions and the ground truths. After this matching, every prediction is associated with a ground truth. Next, we calculate the $l_1$ and $GIoU$ loss on both box and mask predictions and cross entropy loss for class prediction~\cite{carion2020end}. 
\vspace{-0.20cm}
\paragraph{Inference:} The inference pipeline is similar to training pipeline and can be implemented using approximately 50 lines of code. During inference, we first filter the detections with confidences below $50\%$ and a per-pixel max to determine which pixels are predicted to belong to a CH4 plume.



\begin{figure}[t]
\begin{center}
\includegraphics[width=\linewidth]{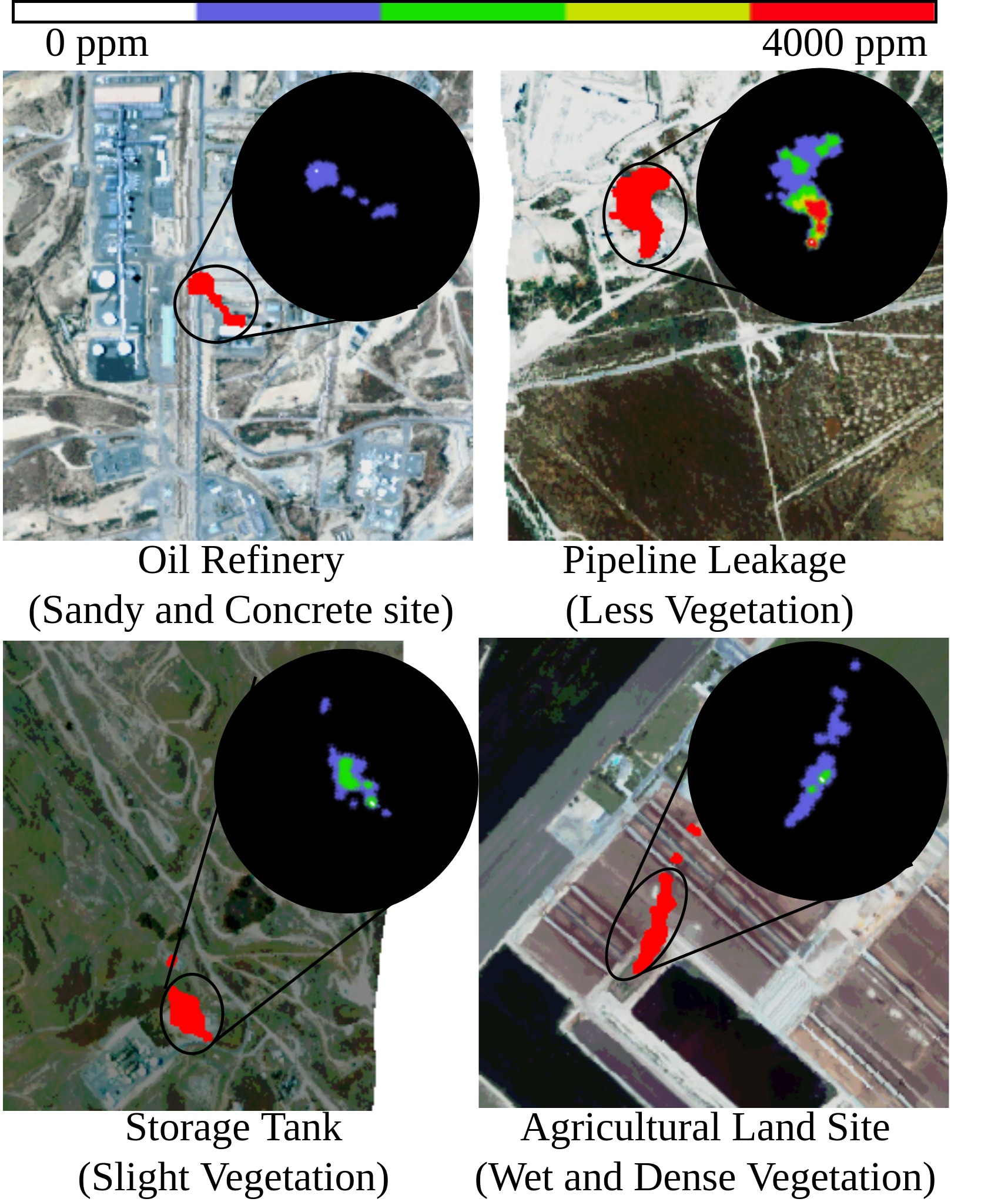}
\end{center}
\vspace{-0.35cm}
   \caption{\it Sample images from MHS dataset. The colormap in black circle shows concentration maps corresponding to the plume mask shown in red. We are showing different types of leakage sources and land cover types. For better visualization, we plotted the binary mask on color image created using visible bands of hyperspectral image.}
\vspace{-0.35cm}
\label{fig:data_sample}
\end{figure}

\section{Methane Hot Spots (MHS) dataset} 
Another significant contribution of  this work is a large scale curated MHS dataset. 
It contains the AVIRIS-NG spectral data with wavelength ranging from $380nm$ to $2510nm$, a $5nm$ sampling~\cite{AVIRIS-NG}, and capturing $432$ channels per pixel. The images from the flight-line are orthorectified and of size $\sim23K \times \sim1.5K \times 432$.
The only currently publicly-available dataset with methane plume segmentation masks is the JPL-CH4-detection-V1.0~\cite{thompson2017isgeo} dataset released by JPL-NASA in 2017. 

The MHS dataset has approximately 4000 plume sites corresponding to approximately 1200 AVIRIS-NG flightlines as shown in Table~\ref{tab:dataset}. MHS also has higher diversity data with flight lines spanning from 2015-2022 and covering terrain from 6 states-- California, Nevada, New Mexico, Colorado, Midland Texas, and Virginia.

\begin{table}
\centering
\begin{tabular}{lcc}
\hline
\multicolumn{1}{c}{\textbf{Dataset}} &
  \textbf{\begin{tabular}[c]{@{}c@{}}MHS (Ours) \\ Dataset\end{tabular}} &
  \textbf{\begin{tabular}[c]{@{}c@{}}JPL-CH4\\ detection-V1.0\end{tabular}}~\cite{thompson2017isgeo} \\ \hline
\# \textit{plume sites }    & \textbf{3961}  & 161          \\
\rowcolor[HTML]{EFEFEF} 
\# \textit{flightlines}     & \textbf{1185}   & 46           \\
\# \textit{point source }   & \textbf{~3675} & ~114          \\
\rowcolor[HTML]{EFEFEF} 
\# \textit{diffused source} & \textbf{~286}  & ~57          \\
\textit{Time period }&
  \textbf{\begin{tabular}[c]{@{}c@{}}2015 - 2022 \\ (~8 years)\end{tabular}} &
  \begin{tabular}[c]{@{}c@{}}2015 \\ (~1 year)\end{tabular} \\
\rowcolor[HTML]{EFEFEF} 
\textit{Segmentation Mask} & \textbf{Yes}   & \textbf{Yes} \\
\textit{Bonding box}       & \textbf{Yes}   & No           \\
\rowcolor[HTML]{EFEFEF} 
\textit{Concentration map} & \textbf{Yes}   & No           \\
\textit{Number of Regions} & \textbf{6}   & 1           \\ \hline
\end{tabular}
\caption{\it Statistics shows MHS dataset comparison with JPL-CH4-detection-V1.0~\cite{thompson2017isgeo} dataset. Each flightline have multiple large and small plume sites. Each flightline have atleast 4 plume sites. The \textit{Point Source} represents high concentration ($300 kg/hr$) to leakage from sources like pipeline leak, storage tanks, oil and gas refineries. \textit{Diffused Source} represent low concentration leakages from sources like biomass degradation in landfills. Our dataset is covers more diverse type of terrain over 6 states.}
\label{tab:dataset}
\end{table}
\vspace{-0.2cm}
\paragraph{Data Pruning:} We selected AVIRIS-NG  flight lines over varying regions as it covers a wide variety of CH$_4$ plume sources, such as leaks in oil and gas refineries, oil and gas extraction points, natural seeps, leaking underground storage tank, coal mines, dairy farms, landfill sites, and pipeline leaks. Along with varying emission sources, we selected regions with  different types of ground terrains like, bare soil, rocks, mountains, light vegetation, water bodies and dense vegetation as shown with few samples in Fig.~\ref{fig:data_sample}. Different types of ground terrain exhibit widely varying albedo and thus have a major impact on the quality of CH$_4$ detections as shown in Fig.~\ref{fig:qual_slf}. Given this, training models with diverse ground terrain data leads to a more robust model.

\subsection{Concentration map and Segmentation mask}
\vspace{-0.15cm}
\noindent \textbf{Concentration map} is provided in the form of a matrix of spatial dimensions same as the  flightline ($\sim23k \times \sim1.5k \times 1$). There is one concentration map per flight-line (orthorectified). It shows methane concentration in parts-per-million (ppm) per-pixel on the ground. Pixel-regions with no methane presence are set to zero. 

\noindent \textbf{Segmentation mask} provided in the format of a \textit{``png"} image file with three channels and of the same spatial dimension as the corresponding flight line ($\sim23k \times \sim1.5k \times 3$).
The segmentation mask is obtained from the concentration mask file by setting all pixel values above zero to represent methane plumes.
We manually annotated \textit{Point Source} and \textit{Diffused Source} based on the type of ground terrain and concentration of methane gas. Following the benchmark dataset~\cite{thompson2017isgeo}, three channels are used to color code \textit{Point Source} (Red) and \textit{Diffused Source} (Green).
The distinction of \textit{Point Source} and \textit{Diffused Source} is derived from the JPL-CH4-detection-V1.0 benchmark dataset~\cite{thompson2017isgeo}. Our annotation style is also consistent with the JPL-CH4-detection-V1.0 benchmark dataset~\cite{thompson2017isgeo}, so that both datasets can be merged seamlessly. 

\subsubsection{Constructing Concentration map}
\label{sec:annot}
Concentration maps are generated by mapping expert-annotated methane-plume concentration maps to the ortho-corrected AVIRIS-NG flightlines. These methane plume annotations are systematically collected from a non-profit~\cite{carbon_mapper} entity. They provide concentration masks of methane emissions in $150 \times 150$ size patches along with location information from different sources (airborne sensors~\cite{gao_sensor}, satellites~\cite{phiri2020sentinel}).
In order to map these patches from different sources to the AVIRIS-NG flight-lines, we use the pixel coordinate locations provided for both the annotations and flight-lines. We use this information to create a homography transformation to map each pixel to its corresponding location in the flight-line. Fig.~\ref{fig:data_sample} shows a sample of varying types of terrains with CH$_4$ segmentation mask in red and concentration mask in black circle.
Details about matching the resolution, ortho-correction, and transformation are discussed in supplementary materials. 
The patch annotations are verified by experts visiting the physical location of emission the same day~\cite{carb}. Most of the regions in California are verified by physical visits by California Air Resource Board~\cite{carbon_mapper, carb}.

\subsection{MHS Statistics}
\vspace{-0.15cm}
MHS statistics and properties are summarized in Table~\ref{tab:dataset}.

\noindent \textbf{Annotations:} MHS provides both segmentation masks and concentration maps which enable development of deep learning algorithms than can produce both CH$_4$ plume location and concentration predictions.

\noindent \textbf{Diversity:} MHS dataset includes AVARIS-NG flightlines spanning 8 years (2015 - 2022) from six states in the U.S.: California, Nevada, New Mexico, Colorado, Texas, and Virginia.


\noindent \textbf{Data Split:} We divide MHS dataset into train/test splits of 80-20\% with overlapping time periods and locations. 
Our dataset covers 6 states.
Each state has sub-regions/locations (e.g. Permian basin) that are covered by multiple non-overlapping flightlines ($25k\times1.5k\times432$ pixels). These flightlines are split into train and test sets. In each set, we create patches ($256\times256\times432$ pixels) from the corresponding flightlines. From the patches/tiles, we take all positives patches (methane (CH$_4$)) and randomly sample equal number of negative (no-CH$_4$) patches. This is done for both train and test sets separately to balance the data and we refer to Section~\ref{sec:ablation} for detailed ablation studies.

\section{Experimental settings}
\vspace{-0.15cm}
\noindent \textbf{Evaluation Metrics}: 
Following the evaluation protocol of H-mrcnn~\cite{kumar2020deep} we report our performance in mean intersection-over-union (mIOU).
Here, mIOU indicates the overlap between the predicted and the ground truth CH$_4$ plume masks. ED represents the accuracy in plume core prediction. Additionally, as first stage of our two stage training procedure contains bounding box prediction, we also report our performance in predicting plume bounding boxes in terms of mean Average Precision (mAP) which tells us the effectiveness of MethaneMapper in eliminating the false positives in plume prediction.  

\noindent \textbf{Data Pre-Processing}: Each input hyperspectral image is approximately of size $25000 \times 1500 \times 432$ taking up memory space of $55-60$ GB. We create tiles of each image in spatial domain, each tile is of size $256\times256\times432$~\cite{kumar2020deep} with an overlap of $128$. The CH$_4$ plume is available in very few pixels in the whole image, $90\%$ of the tiles are negative samples (no methane, just ground terrain). 
We can not use the whole hyperspectral image because of GPU memory limitations

\noindent \textbf{Implementation Details}: The band-selectors module takes 432-channels hyperspectral image as input, the RGB band-selector picks 60 channel from $400nm-700nm$ wavelength range and creates a 3-channel RGB image, the SWIR band-selector picks 100 channel from wavelength range $2000nm-2500nm$. These input images are passed to two ResNet-50~\cite{he2016deep} feature extractor backbones. The backbone networks are initialized with DETR~\cite{carion2020end} trained on COCO dataset~\cite{lin2014microsoft} and input layer initialized randomly~\cite{he2015delving}. The transformer encoder-decoder and our query refiner have 6 layers and 8 heads. We initialized the transformer encoder-decoder with weights extracted and stripped from DETR~\cite{carion2020end} model. The dimension of transformer architecture is $256$ and number of queries is $100$. The \textbf{SFG} module takes in all 432-channels hyperspectral image and generates 1-channel output map of same spatial dimension as input. The feature extractor in \textbf{SFG} is ResNet-50~\cite{he2016deep} initialized with DETR~\cite{carion2020end} trained on COCO dataset~\cite{lin2014microsoft}. The decoder output embeddings are of size $512$. The feature pyramid network in mask prediction module has 3 layers. More details are mentioned in supplementary materials.

\section{Results}
\vspace{-0.15cm}
In this section we will discuss and validate all the design choices for MethaneMapper (MM) with ablations. We show that MM achieves state-of-the-art results in overall performance compared all other methods shown in Tables~\ref{tab:res_main} \&~\ref{tab:res_baseline}.

\begin{table}
\centering
\adjustbox{width=\columnwidth}{
\begin{tabular}{lllllll}
\hline
 &
  Methods &
  \cellcolor[HTML]{EFEFEF}\begin{tabular}[c]{@{}l@{}}Back\\ bone\end{tabular} &
  \begin{tabular}[c]{@{}l@{}}SFG\\ F.Ext.\end{tabular} &
  \cellcolor[HTML]{EFEFEF}\#params &
  mAP &
  \cellcolor[HTML]{EFEFEF}mIOU \\ \hline
 &
  \multicolumn{6}{c}{\textit{JPL-CH4-detection-v1.0 Dataset}} \\ \hline
\multicolumn{1}{l|}{1} &
  Hu et. al &
  \cellcolor[HTML]{EFEFEF}R-50 &
  - &
  \cellcolor[HTML]{EFEFEF}75M &
  0.26 &
  \cellcolor[HTML]{EFEFEF}0.48 \\
\multicolumn{1}{l|}{2} &
  H-mrcnn &
  \cellcolor[HTML]{EFEFEF}R-50 &
  - &
  \cellcolor[HTML]{EFEFEF}353M &
  0.53 &
  \cellcolor[HTML]{EFEFEF}0.86 \\
\multicolumn{1}{l|}{3} &
  MM &
  \cellcolor[HTML]{EFEFEF}R-50 &
  R-50 &
  \cellcolor[HTML]{EFEFEF}\textbf{80M} &
  \textbf{0.63} &
  \cellcolor[HTML]{EFEFEF}\textbf{0.91} \\ \hline
 &
  \multicolumn{6}{c}{\textit{MHS (Ours) Dataset}} \\ \hline
\multicolumn{1}{l|}{4} &
  SpectralFormer &
  \cellcolor[HTML]{EFEFEF}R-50 &
  - &
  \cellcolor[HTML]{EFEFEF}84M &
  0.33 &
  \cellcolor[HTML]{EFEFEF}0.41 \\
\multicolumn{1}{l|}{5} &
  UPSNet (stuff) &
  \cellcolor[HTML]{EFEFEF}R-50 &
  - &
  \cellcolor[HTML]{EFEFEF}69M &
  0.32 &
  \cellcolor[HTML]{EFEFEF}0.38 \\
\multicolumn{1}{l|}{6} &
  \begin{tabular}[c]{@{}l@{}}UPSNet (stuff\\ + things)\end{tabular} &
  \cellcolor[HTML]{EFEFEF}R-50 &
  - &
  \cellcolor[HTML]{EFEFEF}69M &
  0.29 &
  \cellcolor[HTML]{EFEFEF}0.35 \\
\multicolumn{1}{l|}{7} &
  DETR &
  \cellcolor[HTML]{EFEFEF}R-18 &
  * &
  \cellcolor[HTML]{EFEFEF}33M &
  0.37 &
  \cellcolor[HTML]{EFEFEF}0.56 \\
\multicolumn{1}{l|}{8} &
  DETR &
  \cellcolor[HTML]{EFEFEF}R-50 &
  * &
  \cellcolor[HTML]{EFEFEF}59M &
  0.44 &
  \cellcolor[HTML]{EFEFEF}0.59 \\ \hline
\multicolumn{1}{l|}{10} &
   &
  \cellcolor[HTML]{EFEFEF}R-18 &
  \begin{tabular}[c]{@{}l@{}}Linear\\ Layer\end{tabular} &
  \cellcolor[HTML]{EFEFEF}39M &
  0.45 &
  \cellcolor[HTML]{EFEFEF}0.60 \\
\multicolumn{1}{l|}{11} &
   &
  \cellcolor[HTML]{EFEFEF}R-18 &
  R-18 &
  \cellcolor[HTML]{EFEFEF}44M &
  0.52 &
  \cellcolor[HTML]{EFEFEF}0.63 \\
\multicolumn{1}{l|}{12} &
  \multirow{-3}{*}{MM} &
  \cellcolor[HTML]{EFEFEF}R-50 &
  R-50 &
  \cellcolor[HTML]{EFEFEF}\textbf{80M} &
  \textbf{0.59} &
  \cellcolor[HTML]{EFEFEF}\textbf{0.68} \\ \hline
\end{tabular}
}
\caption{\it Comparison with baselines. ``-" represent Not Applicable and ``*" represent no \textbf{SFG} module and a random query used for transformer decoder. The top section shows performance on JPL-CH$_4$ dataset~\cite{thompson2017isgeo}. MethaneMapper achieves better results than heavily tuned H-mrcnn with $\sim 5\times$ fewer parameters. The overall detection accuracy is higher on this dataset because the type of ground terrain is uniform across all flightlines.
In MHS dataset, MM outperforms multiple baselines as shown in rows 4-12. MM accuracy is lower in MHS than JPL-CH$_4$ dataset because MHS dataset has more variety of ground terrain spreading over 6 states}
\label{tab:res_main}
\end{table}

\begin{table}
\centering
\adjustbox{width=0.6\linewidth}{
\begin{tabular}{lll}
\hline
Methods                       & mAP           & mIOU          \\ \hline
LogReg~\cite{cheng2006logistic}                        & -             & 0.05          \\
SVM~\cite{shah2007ica}                           & -             & 0.29          \\
PCA + LogReg                  & -             & 0.06          \\
PCA + SVM                     & -             & 0.31          \\ \hline
\rowcolor[HTML]{EFEFEF} 
\textbf{MM (R-50)} & \textbf{0.63} & \textbf{0.91} \\ \hline
\end{tabular}
}
\caption{\it Comparison with classical machine learning methods. ``-" represent Not Available. The classical ML methods are not suited for the CH$_4$ detection task. MethaneMapper outperforms all methods on JPL dataset~\cite{thompson2017isgeo}}
\vspace{-0.7cm}
\label{tab:res_baseline}
\end{table}

\subsection{Performance comparison}
\vspace{-0.15cm}
\noindent \textbf{Deep Learning methods}: We trained MM with ResNet-50~\cite{he2016deep} backbone on the same dataset that H-mrcnn~\cite{kumar2020deep} (JPL-CH4-detection-V1.0~\cite{thompson2017isgeo}) was trained on for fair comparison. To align with H-mrcnn we used the same split and input image size. 
The MM model with $80$M parameters trained for 250 epochs outperforms by significant margin the H-mrcnn model with $352$M parameters. Results are summarized in Table~\ref{tab:res_main} that includes the performance of MM on the new larger MHS dataset. We note that though the code for H-mrcnn is available, many of the modules are deprecated and can not be reproduced. 
The 'Backbone' column represents backbones used for feature extraction from input image,'SFG F.Ext.' represents the feature extractor in \textbf{SFG} module in MethaneMapper.
We observed (qualitatively) that H-mrcnn fails to detect small CH$_4$ plumes with concentration lower than 100kg/hr while MM detects those. 

We did evaluation by implementing 3 baseline models~\cite{hong2021spectralformer, carion2020end, xiong2019upsnet} shown rows 4-8 of Table~\ref{tab:res_main}. These methods were not designed for CH$_4$ detection task, therefore we needed to modify their input channel size. The poor performance of these methods may be attributed to the weak signal of interest in a high dimensional data, high number of confusers, and limited annotated data. Additionally, the only hyperspectral baseline method SpectralFormer~\cite{hong2021spectralformer} has low efficiency due its pixel-wise training scheme.

\noindent \textbf{Classical ML methods:} We trained and tested multiple existing machine learning based approaches that are used for methane detection, performance shown in Table~\ref{tab:res_baseline}. Logistic regression (LogReg)~\cite{cheng2006logistic} and multinomial logistic regression (MLR)~\cite{khodadadzadeh2014subspace} failed to produce any meaningful detection with $90\%$ false positive detections. 
We also trained a Support Vector Machine (SVM) ~\cite{shah2007ica, tax2004support} based classifier, it performed slightly better than LR and MLR methods with an IOU of 21\%. SVMs are prone to false positives detections same as Gaussian Mixture Models~\cite{shah2007ica}. We observed that all traditional methods are not suited for the task of CH$_4$ detection. We also tested reducing the dimension using principal component analysis (PCA) or just taking bands which shows maximum CH$_4$ absorption. In the later case, the traditional methods performed better than using all 432 bands, this backs our idea of just using bands from SWIR region. 

\noindent \textbf{Qualitative results.} Fig.~\ref{fig:qual_res} shows comparison of MM's mask and bounding box prediction with ground truth mask on different ground terrains. The Leakages are from different type of sources such as, oil refinery, pipeline and storage tank. MM makes correct predictions in varying scenarios.

\subsection{Ablation Studies}
\label{sec:ablation}
\vspace{-0.15cm}
We did the experiments for ablation on MHS dataset with ResNet-50 as backbone and validate the design choices. One parameter is changed for each ablation and others kept at best settings. More ablations in Supplementary.

\begin{figure}[t]
\begin{center}
\includegraphics[width=1\linewidth]{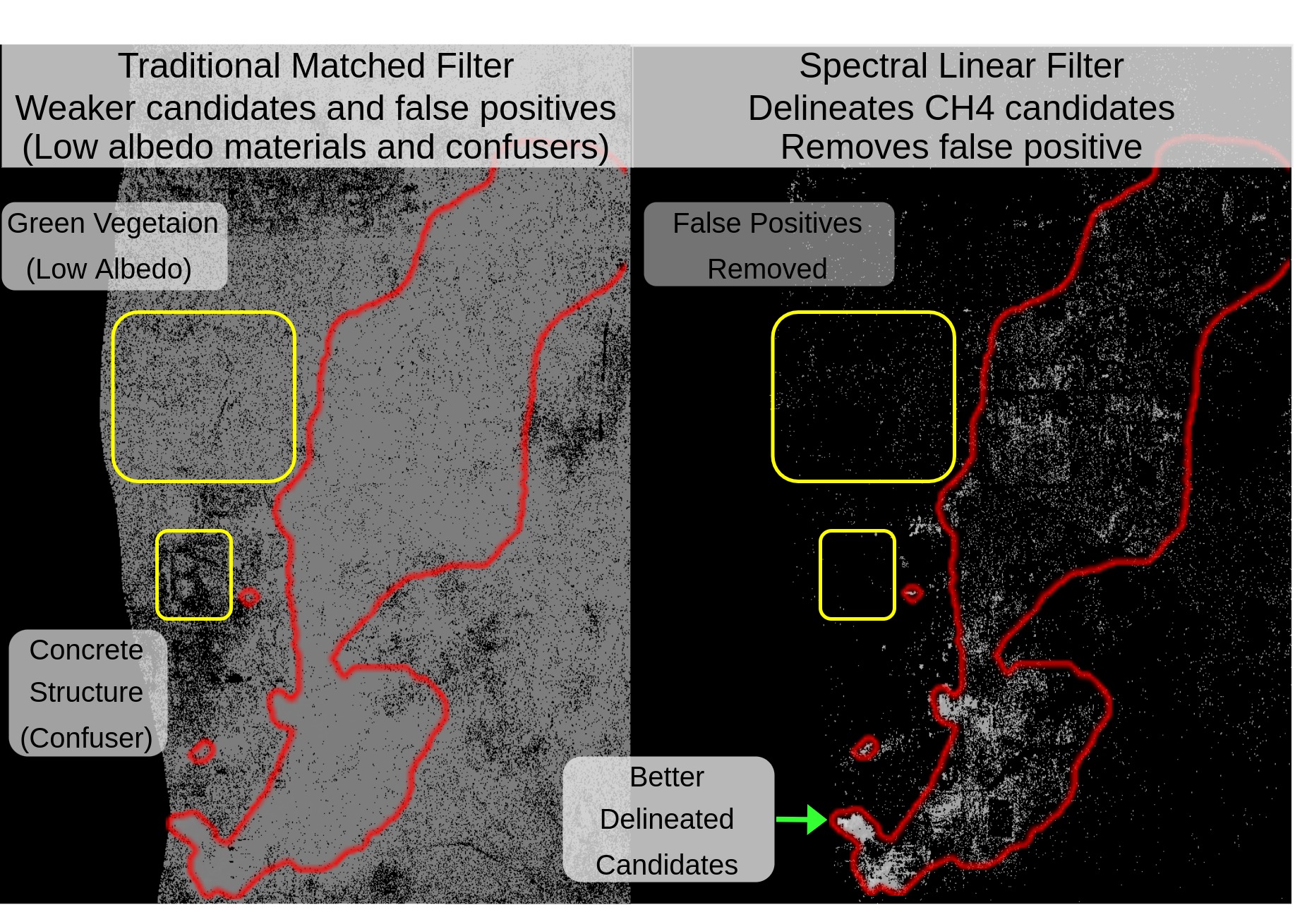}
\end{center}
\vspace{-0.75cm}
   \caption{\it Comparison of SLF with traditional filter in SFG module. White pixels represent methane and black no-methane. Red boundary represents ground-truth plume mask. SLF module generates better CH$_4$ candidates}
\vspace{-0.45cm}
\label{fig:qual_slf}
\end{figure}

\vspace{-0.15cm}
\paragraph{Spectral Feature Generator Module:} In Table~\ref{tab:res_main} lower section, we show the effectiveness of our \textbf{SFG} module for the query refiner block. Our baseline is standard implementation of DETR~\cite{carion2020end} for segmentation task represented by row-1 and row-2 of Tab.~\ref{tab:res_main} lower section. Using CH$_4$ candidates feature from \textbf{SFG} improves the bounding box detection performance by 0.14 mAP and mask prediction by 0.09 mIOU. This demonstrate that guiding queries with CH$_4$ candidates feature generated by \textbf{SFG} produces better embeddings as compared to random queries.

Along with this, we explored the provision of CH$_4$ candidates feature at 2 places, (i) at input level concatenating it with $f_{comb}$; and (ii) as input to query refiner. We see an improvement of 0.09 mAP and 0.08 mIOU when \textbf{SFG} module output is passed to query refiner. We hypothesize that this is because on concatenating with input features, the CH$_4$ candidates feature information gets lost, while as cross-attention with queries reduces the search space for decoder and generate better embeddings. 

We also experimented with different types of feature extractors for \textbf{SFG} module, and observed that a Resnet18 or Resnet50~\cite{he2016deep} is more effective than a 2 linear layer feature extractor as shown in Table~\ref{tab:res_main}.

\begin{figure}[t]
\begin{center}
\includegraphics[width=0.8\linewidth]{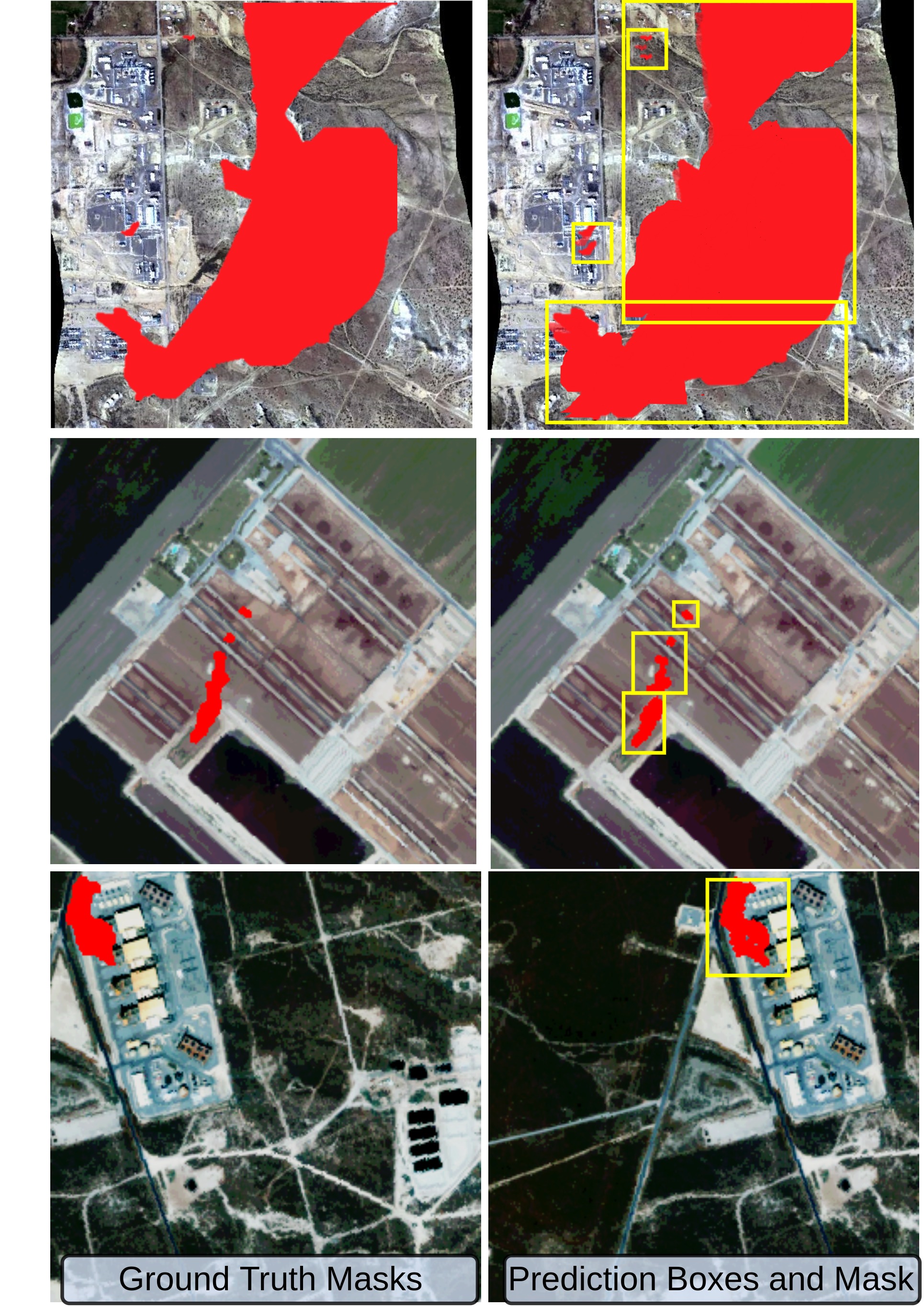}
\end{center}
\vspace{-0.65cm}
   \caption{\it Sample ground truths and predictions on MHS dataset. We show robustness of MethaneMapper predictions on different kind of ground terrain, rows 1 and 3 shows leakage at a refinery, row 2 shows leakage from pipeline in agricultural land, row 4 shows leakage from storage tank with concrete background.}
\vspace{-0.35cm}
\label{fig:qual_res}
\end{figure}

\noindent \textbf{Spectral Linear Filter:} We  experimented with SLF for computing covariance (\textit{Cov}) using different subset of columns in the input hyperspectral image. We observed that the SLF is most effective when covariance is computed class-wise based on land cover. Class-wise \textit{Cov} ensures that the radiance absorption by ground terrain is same for all the pixels while computing CH$_4$ enhancement. As can be seen in Fig.~\ref{fig:qual_slf}, {SLF} amplifies CH$_4$ candidate detection and reduces false positives. {SLF} leads to a 0.03 mAP improved in detection compared to traditional filters. The prediction from MM is shown row-1 of Fig.~\ref{fig:qual_res}.

\noindent \textbf{Geographic generalization:} To assess the geographical generalization capabilities of MM, we trained it on MHS data from all states except California and tested it on flightlines from California. We observed a slight drop of 0.04 mAP in detections. However, when trained on all data except Virginia, we noticed a significant drop of 0.09 mAP in detections. We attribute this to the fact that the land cover in Virginia is dense and moist vegetation, has a lower solar reflectance compared to the arid regions of California, Texas, and Nevada. 

\noindent \textbf{Temporal generalization:} Testing MM on 2015 after training on data from 2016-2022 showed no performance drop.

\noindent \textbf{Unbalanced test set:} MM's performance dropped by 0.05 mAP on an unbalanced test set with only 10\% positive samples (CH$_4$) and 90\% negative samples (no-CH$_4$). This highlights the challenges in CH$_4$ detection. Future work will address this issue.

\vspace{-0.2cm}
\section{Conclusion}
\vspace{-0.15cm}
This paper presents MethaneMapper --  a hyperspectral Transformer for methane plume detection. It utilize spectral and spatial correlations using a spectral feature generator and a query refiner, to accurately delineate the CH$_4$ plumes. Additionally, we curated a large-scale dataset for the task, a first of its kind, which will be made available to all researchers. The proposed MethaneMapper significantly improves upon the current methods in terms of detection and localization accuracy, as our extensive experiments demonstrate. Future work will extend the  model to global monitoring~\cite{kumarguided} using multispectral satellite imaging data.

\section{Acknowledgments}
\vspace{-0.15cm}
This research is partially supported by the following grants: NSF award SI2-SSI \#1664172 and US Army Research Laboratory (ARL) under agreement number W911NF2020157.

{\small
\bibliographystyle{ieee_fullname}
\bibliography{egbib}
}

\clearpage
\thispagestyle{empty}

\section{SUPPLEMENTARY MATERIALS}
In supplementary section, we provide will all the details about the data collection and annotations creation process. We also provide with the complete derivation of Spectral Linear Filter (\textbf{SLF}) along with a pseudo implementation of \textbf{SLF} algorithm. Next in the document we provide some more qualitative examples of success and failure cases of MethaneMapper. Towards the end of the document we provide graph plots about training convergence of all the ablation experiments with Spectral Feature Generator (SFG) and Query Refiner (QR) module.

\subsection{Dataset}
\subsubsection{AVIRIG-NG} AVIRIS-NG~\cite{AVIRIS-NG} is an acronym for the \textit{Airborne Visible InfraRed Imaging Spectrometer - Next Generation} developed by Jet Propulsion Laboratory (JPL) in 2009. JPL conducted thousands of flight lines recording data with AVIRIS-NG instrument in last 7 years. On the AVIRIS-NG instrument an array of total $598$ sensors in push-broom order captures an unortho-rectified data-cube of spatial dimension $\sim23k \times 598$, where each sensor records a spectral wavelengths ranging from $380nm - 2510nm$ ~\cite{5747395} making a dimension of $432$ channels.
It has $34^o$ field of view with a 1 mrad instantaneous field of view the generates spatial resolution of $1-8m$ based on altitude.
This data is then rectified using a geometric lookup table and the resulting data cube is of size $\sim23k \times \sim1.5k \times 432$.
The data is provided in Band Interleaved by Line (BIL) ordering. BIL ordering signifies the 3D matrix is indexed first by image row, then by channel, and then by the image column~\cite{thompson2017isgeo}.
One can find details about the naming convention and the type of data each files contain in ``README.txt" file in each flightline folder.
The data can be loaded into a $numpy$ array easily using python libraries. All data is orthorectified.

\subsubsection{Annotations}
\begin{figure}[t]
\begin{center}
\includegraphics[width=\linewidth]{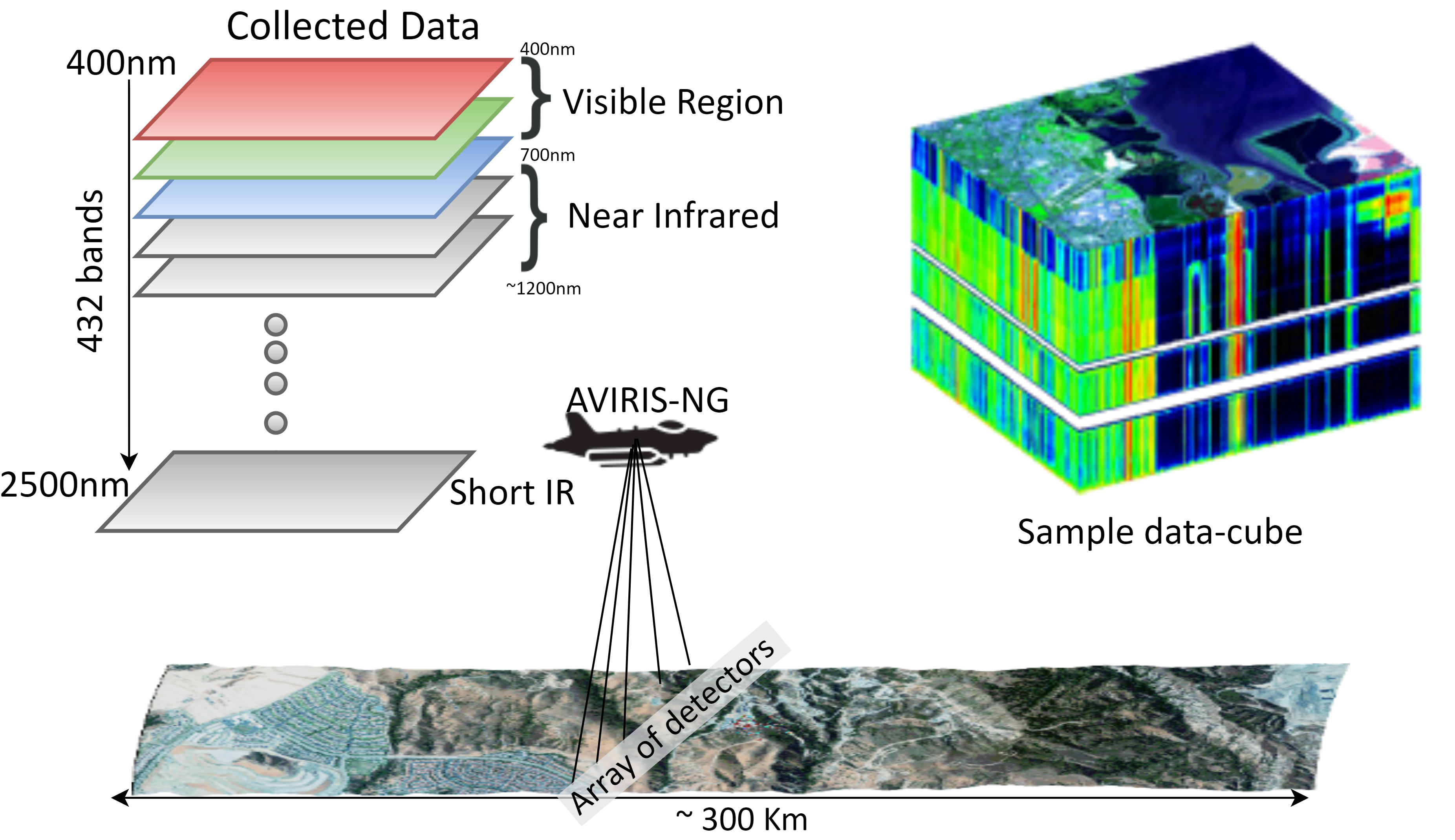}
\end{center}
\vspace{-0.45cm}
   \caption{\it Depiction of data collection process. Each flightline is $\sim300$ kms long. An array of 598 sensors records data at 1.5m/pixel spatial resolution. All flightlines are ortho-corrected. Each data-cube is of dimension $\sim23k \times \sim1.5k \times 432$.}
\label{fig:data_cube2}
\end{figure}
\begin{figure}[t]
\begin{center}
\includegraphics[width=1\linewidth]{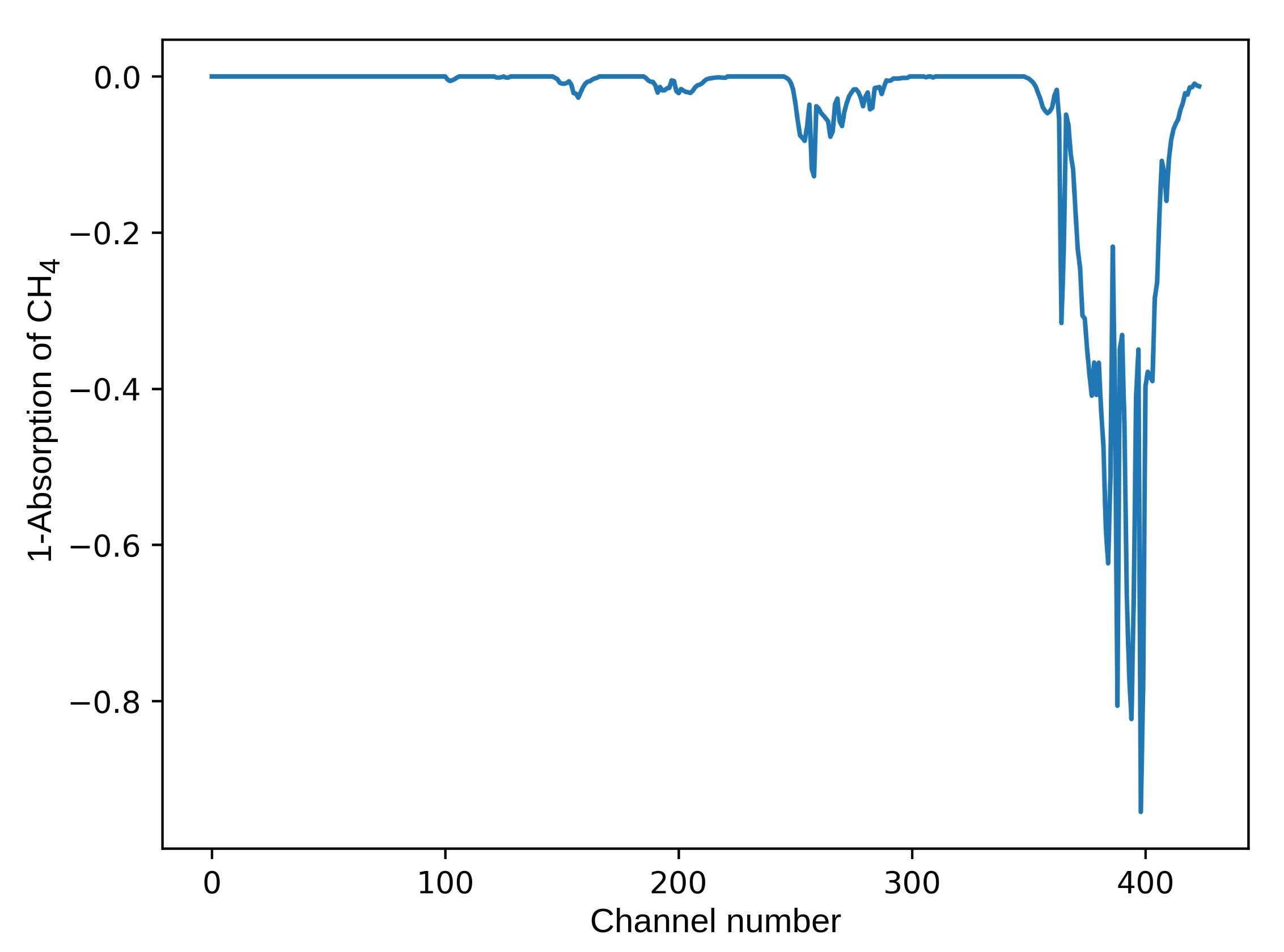}
\end{center}
\vspace{-0.45cm}
   \caption{\it Spectral absorption pattern of CH$_4$ gas. The x-axis show the channel number ranging from 0-400 corresponding to wavelength range ($400nm-2500nm$). It is obtained from the public repository HITRAN~\cite{gordon2022hitran2020}.}
\label{fig:absorb}
\end{figure}

\noindent\textbf{Transformation and Ortho-correction.} First step is to read the annotation GeoTiff patch of size $150\times150$ of a methane concentration mask and convert its Coordinate Reference System (CRT) to AVIRIS-NG flightlines' CRT (EPSG 4326). 
Next, we use the corresponding AVIRIS-NG flightlines' geometric lookup table and unortho-corrected geographic pixel location to generate ortho-corrected geographic pixel location data of the flightline.
Next, we find the flightline's geographic indices that are closest to the geographic indexes of the methane concentration mask (annotation GeoTiff). 
Finally, we use these corresponding pixels to compute a homography transform matrix that maps the methane concentration mask (annotation GeoTiff) to the AVIRIS-NG flightline's spatial dimensions. We repeat this process for each plume in the flightline in order to generate the CH$_4$ concentration map for the entire flightline.

\noindent \textbf{Resolution matching.} To match the resolution of transformed annotation GeoTiff patch to AVIRIS-NG flightline, we use nearest-neighbor resampling. A pixel from the transformed annotation GeoTiff patch may be repeated multiple times in the CH$_4$ concentration map for the entire flightline.

\noindent \textbf{Annotation Style.}
The \textit{Point Source} and \textit{Diffused Source} are coded following the same standard as JPL-CH4-detection-V1.0~\cite{thompson2017isgeo} dataset. The 3-channels have values in [0-255] range.
\begin{itemize}
    \item Red (255,0,0): plume, believed to be associated with a \textit{Point Source}
    \item Blue (0,0,255): plume, believed to be associated with a \textit{Diffuse Source}
    \item Black (0,0,0): no plume (or unlabeled)
\end{itemize}
We kept our annotation style consistent with JPL-CH4-detection-V1.0 benchmark dataset~\cite{thompson2017isgeo} so that both JPL-CH4-detection-V1.0 and MHS datasets can be merged seamlessly.

\begin{figure*}[t]
\begin{center}
\includegraphics[width=0.7\linewidth]{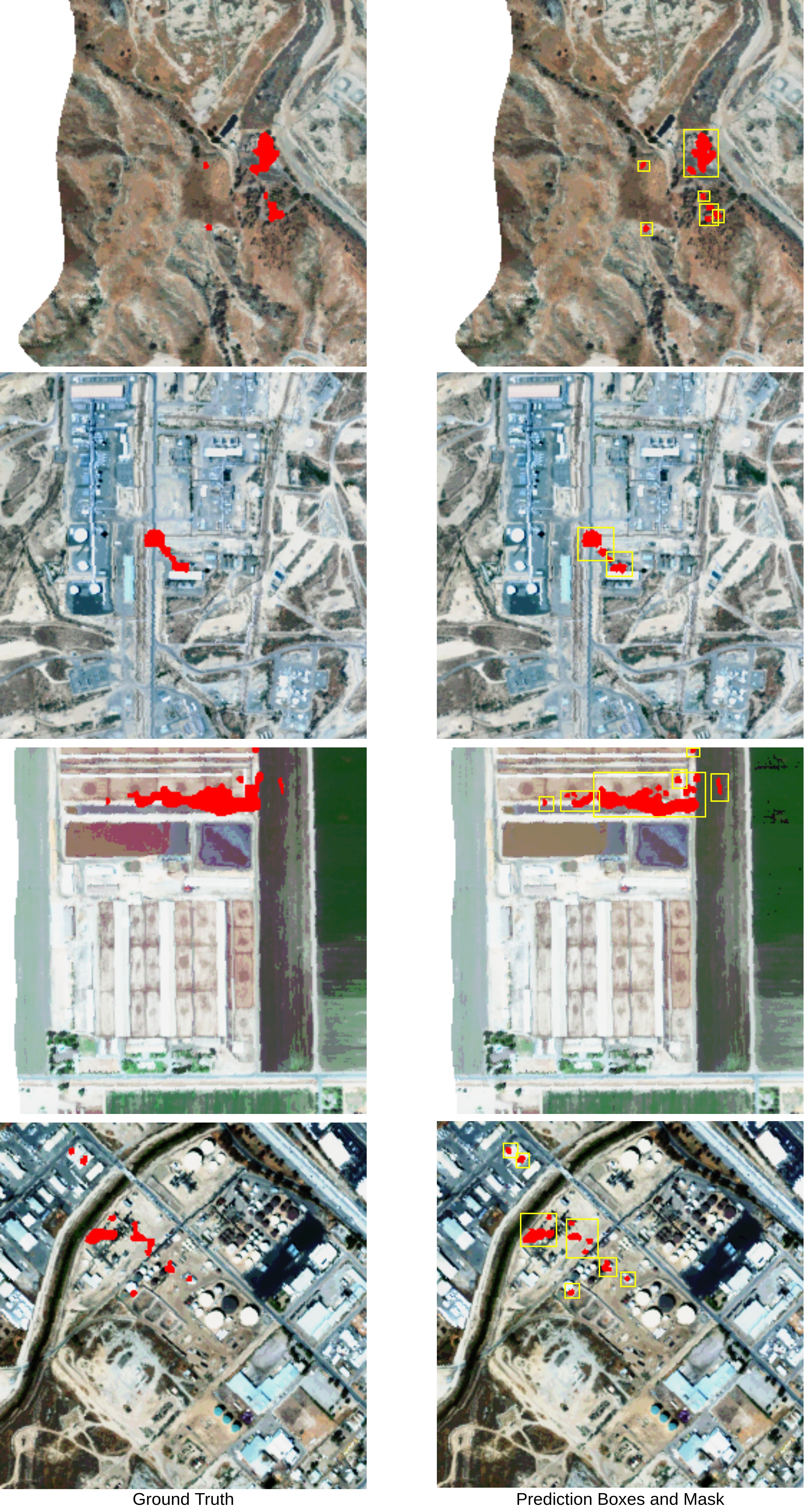}
\end{center}
\vspace{-0.45cm}
   \caption{\it Sample ground truths and predictions on MHS dataset. We are showing different type of terrains and CH$_4$ predictions on them. The type of emission source in all samples varies too.}
\label{fig:qual_res1}
\end{figure*}

\subsection{Spectral Linear Filter(SFL)}

\subsubsection{Traditional Matched Filter}
Passive hyperspectral imaging sensors captures spectral radiances values from $N_0$ ($N_0 = 432$) spectral channels corresponding to wavelengths ranging from $400nm-2500nm$ as shown in Fig.~\ref{fig:data_cube2} with sample data-cube. The complete hyperspectral image is represented as \textbf{x} $\in \mathbb{R}^{H_0 \times W_0 \times N_0}$ where $H_0,W_0\& N_0$ are height, width and number of channels respectively. In this hyperspectral data, we are looking for a very weak signature of interest hidden in background. In this case the signature of interest is CH$_4$ and the background is ground terrain. CH$_4$ shows strong absorption patterns around $2100nm-2500nm$ wavelength.

The most common linear approach for finding CH$_4$ candidates is taking a $N_0$-dimension (same as number of spectral channels) vector $\alpha$, and apply as a dot product to each pixel ($N_0$-dimension) in the hyperspectral image to generate a scalar output per pixel. This operation is supposed to reduce or remove the ground terrain, sensor noise and amplifies CH$_4$ signature. The $\alpha$ vector used here is called as ``matched filter". Therefore computing right $\alpha$ is very critical for generating better candidates of CH$_4$ emission. It is dependent on absorption pattern of CH$_4$ and on the distribution of the ground terrain. 
To model $\alpha$, let $\textbf{r}_i \in\ \mathbb{R}^n$ be a $i^{th}$ pixel from the hyperspectral image representing the ground terrain pixel and sensor noise, and $\textbf{t}$ be the CH$_4$ absorption pattern~\cite{gordon2022hitran2020}. 
This is modeled as the additive perturbation as shown below:
\begin{equation}
    \textbf{x}_i = \textbf{r}_i + \textbf{t},
\end{equation}
where $\textbf{x}_i$ is the spectrum when CH$_4$ is present.
The CH$_4$ absorption pattern \textbf{t} represents the change in radiance units of the background caused by adding a unit mixing ratio length of CH$_4$ absorption~\cite{funk2001clustering, kumar2020deep}. Figure~\ref{fig:absorb} shows the spectral absorption pattern of CH$_4$ per channel.
In the ideal scenario where only CH$_4$ gas is present in signal (i.e. all white background), the matched filter output is $\alpha^T\textbf{t}$. In case there is no gas and just ground terrain and sensor noise, the matched filter output is $\alpha^T\textbf{r}_i$. The variance ($Var$) of $\alpha^T\textbf{r}_i$ for latter is represented by: 
\begin{equation}
    Var(\alpha^T\textbf{r}_i) = \langle(\alpha^T\textbf{r}_i-\alpha^T\boldsymbol{\mu})^2\rangle = \alpha^T\textbf{Cov}\alpha,
\end{equation}
where $\textbf{Cov}$ and $\mu$ are covariance and mean respectively computed for $\textbf{r}_i$. Inspired from~\cite{kumar2020deep, funk2001clustering} we define the Methane-to-Ground terrain Ratio (MGR) is:
\begin{equation}
    \text{MGR} = \frac{|\alpha^T\textbf{t}|^2}{\alpha^T\textbf{Cov}\alpha},
\end{equation}
We can see that the magnitude of $\alpha$ does not affect MGR. According to~\cite{theiler2007beyond, funk2001clustering, kumar2020deep}, the MGR can be maximized subject to constraints(zero mean and $\alpha^T\textbf{K}\alpha$ constraint to 1). The matched filter $\alpha$ is then represented by:
\begin{equation}
    \alpha = \frac{\textbf{Cov}^{-1}\textbf{t}}{\sqrt{\textbf{t}^T\textbf{Cov}^{-1}\textbf{t}}}.
\end{equation}
In ideal instances when there is no background (i.e. all white background) and just CH$_4$ gas present. The matched filter in equation~\ref{eq:mf} is directly proportional to \textbf{t}. This is just the target signature (\textbf{t}) itself scaled so that the filtered output has variance of one. The methane enhancement per pixel can be computed as follows:

\begin{equation}
    \hat{\alpha}(\textbf{x}_i) = \frac{(\textbf{x}_i-\mu)^T\textbf{Cov}^{-1}\textbf{t}}{\sqrt{\textbf{t}^T\textbf{Cov}^{-1}\textbf{t}}},
    \label{eq:mf}
\end{equation}
where $\hat{\alpha}(\textbf{x}_i)$ is the per pixel estimation of methane, on other words, column enhancement of methane.
The covariance matrix ($\textbf{Cov}$) used is not known as $prior$ and is estimated from data. It is computed as outer product of the mean subtracted radiance over all the pixels. In other words, the traditional matched filter from equation~\ref{eq:mf} computes the covariance ($\textbf{Cov}$) of ground terrain with an underlying assumption that in all elements have similar absorption pattern. Same covariance matrix ($\textbf{Cov}$) matrix is used to whitens the varying ground terrain and amplify the CH$_4$ present.
But in realistic scenarios, the ground terrain is varying, the type of terrain changes frequently, there is water bodies, bare soil, vegetation, dense vegetation, building structures in cities, roads etc in a single image. For example, water have a strong absorption of solar radiations, therefore the methane on such backgrounds have a very weak visibility. Similarly, wet fields  dense vegetation have similar behaviour. On the other hand, bare soil, rocks, etc have lower absorption, the methane present on such background have strong visibility. 
A simple and single approximation of the covariance ($\textbf{Cov}$) of ground distribution can not provide the right and effective estimate of methane enhancement. To tackle this limitation, we developed an spectral linear Filter (SLF) that does land cover classification and segmentation and reduces the noise as discussed in the next sections.

\subsubsection{Landcover Classification and Segmentation}
In this section, we improve upon the limitations mentioned in the previous section. We start with taking hyperspectral bands from visible spectrum ($400 nm - 700 nm$) and near-mid infrared region ($800nm-1350nm$). We recreated the $RGB$ representation of the ground terrain by a weighted normal distribution for each color band. Same is done for near infrared region. Next we take a simple, very effective and efficient approach for doing landcover classification and segmentation. We compute the Normalized Difference Vegetation Index (NDVI)~\cite{pettorelli2013normalized, normalized-difference-vegetation-index} and Normalized Difference Water Index (NDWI)~\cite{gao1996ndwi}. NDVI quantifies vegetation by measuring the difference between near-infrared (which vegetation strongly reflects) and red light (which vegetation absorbs)~\cite{normalized-difference-vegetation-index}. It ranges from $-1$ to $+1$. It is a very effective index and has been used in literature for more than 4 decades.
~\cite{gao1996ndwi} created NDWI and used it to highlight open water features in a satellite image, allowing a water body to “stand out” against the soil and vegetation. It is calculated using the GREEN-NIR (visible green and near-infrared) and ranges from $-1$ to $+1$. Its primary use today is to detect and monitor slight changes in water content of the water bodies. 
\begin{equation}
    ndvi = \frac{NIR - R}{NIR+R};\;\; ndwi = \frac{NIR-MIR}{NIR+MIR}
\end{equation}
where $NIR$ is near infrared region normalized around $880nm$, $MIR$ is mid infrared normalized around $1240nm$ and $R$ is red, normalized around $660nm$. We take advantage of these indexes and create segmentation maps for different types of vegetation, water bodies, bare soil, rocks, mountains, city/urban areas, roads etc. We take the classification thresholds from~\cite{kriegler1969preprocessing, NDVI-classification}. 
For simplification, we also tested by splitting the scale $-1$ to $+1$ in 20 classes, each with a range of $<0.1>$. We obtained comparable results as compared to using classification ranges from~\cite{kriegler1969preprocessing, NDVI-classification}. 
This simple, effective and efficient approach gives three fold boost to our spectral linear filter CH$_4$ candidates estimation. 

\subsubsection{Cov per class} 
We take the segmented image from previous step, we will call segmented image as segmentation mask for simplicity now onward. In practice we have 20 classes, each with a segmentation mask. We merged two or more adjacent classes into one if the number of pixels in that class is less $10000$ . The Number of pixels in each class is kept higher to ensure that while computing the covariance ($\textbf{Cov}$) matrix, the methane signal does not have any or have negligible effect. It is okay to merge adjacent classes into one because they have almost similar radiance/reflectance,  for example, light vegetation and normal vegetation have similar reflectance, etc. For each class we compute a separate mean and covariance matrix. The covariance $\textbf{Cov}_k$ of $k^{th}$ class is computed as:
\begin{equation}
    \textbf{Cov}_k = \frac{1}{N}\sum_{i=1}^{i=j}(\textbf{x}_i - \mu_k)(\textbf{x}_i - \mu_k)^T \; \forall \; j \in k,
\end{equation}
where $N$ is the number of pixels ($>10000$) in $k^{th}$ class and $\mu_k$ is the mean of $k^{th}$ class. For each class we compute the mean $\mu_k$, covariance matrix $\textbf{Cov}_k$. While iterating through each pixel of hyperspectral image, we check to which class $k$ the pixel $\textbf{x}_i$ belongs to and use those pre-computed values. The final Spectral Linear Fitler (\textbf{SLF}) is shown as below:

\begin{equation}
    \textbf{SLF} (\textbf{x}_{i}) = \frac{(\textbf{x}_{i}-\mu_k)^T\textbf{Cov}_k^{-1}\textit{t}}{\sqrt{\textit{t}^T\textbf{Cov}_k^{-1} \textit{t}}} \; \forall\; (i) \in \text{class}~k
\end{equation}

where $\textbf{Cov}^{-1}$ is the inverse of covariance matrix. Next to suppress the sensor noise, we exploit the simple method of tracking each sensor. Each sensor have different physical properties, that can influence the data captured by it. We track each individual sensor in the flight line. Since the data is rectified, the data from each sensor does not belong to single column, instead it is spread randomly across all the columns. This is dependent on the flight path and the movement in the airplane while moving. We used simple data-structure algorithms like depth first search. Tracked each boundary pixels and assigned them to single sensor. We used data from 10-15 adjacent sensor at one time, normalize it and then compute the covariance matrix in previous step with segmentation mask. Our approach is very simple and straight forward.

The algorithm~\ref{algo:mf} shows the pseudo code for our Spectral Linear Filter (\textbf{SLF}).

\begin{algorithm}
{\small
 \KwData{\textit{MHS dataset}}
 \KwResult{\textit{CH$_4$ concentration map}}
 initialization\;
 
 \For{$mhs$ in $MHS$}{
    1. create memory map $mhs$\;
    2. seg\_mask = compute segmentation mask;\\
    \For{mask in seg\_mask}{
        data.\textbf{append}($mhs$[mask])\\
        \textbf{if} (len(data) $< 100000$): \textbf{continue}\\
        \textbf{Cov}, $\mu$ = compute\_stats(data);\\
    }
    3. sensor\_array = individual sensors;\\
    \For{arrays in sensor\_array}{
        data = $mhs$[arrays]\\
        \For{x$_i$ in data}{
            k = seg\_mask[i];\\
            $\textbf{SLF}(\textbf{x}_i) = \frac{(\textbf{x}_i-\mu_k)^T\textbf{Cov}^{-1}_k\textbf{t}}{\sqrt{\textbf{t}^T\textbf{Cov}^{-1}_k\textbf{t}}}$\\
        }
    }
 $\textbf{SLF}(\textbf{x}_i) \; \forall\; classes\; and\; i \in mhs$
 }
 }
 \caption{ Enhanced Matched Filter}
 \label{algo:mf}
\end{algorithm}

\subsubsection{Training policy} We trained MethaneMapper in two styles, (i) pre-training the bounding box and class detection first and then freezing the pre-trained model parameters and training only the mask prediction layer; and (ii) trained whole pipeline end-to-end and achieved similar performance on both the cases. 

\subsubsection{Qualitative Results}
In this section we show few more qualitative examples of CH$_4$ plume mask prediction and few cases where MM failed to detect any CH$_4$ gas emission.Figure.~\ref{fig:qual_res1} shows the CH$_4$ detections in different types of background terrain and different types of emission source. 

\begin{figure*}[t]
\begin{center}
\includegraphics[width=1\linewidth]{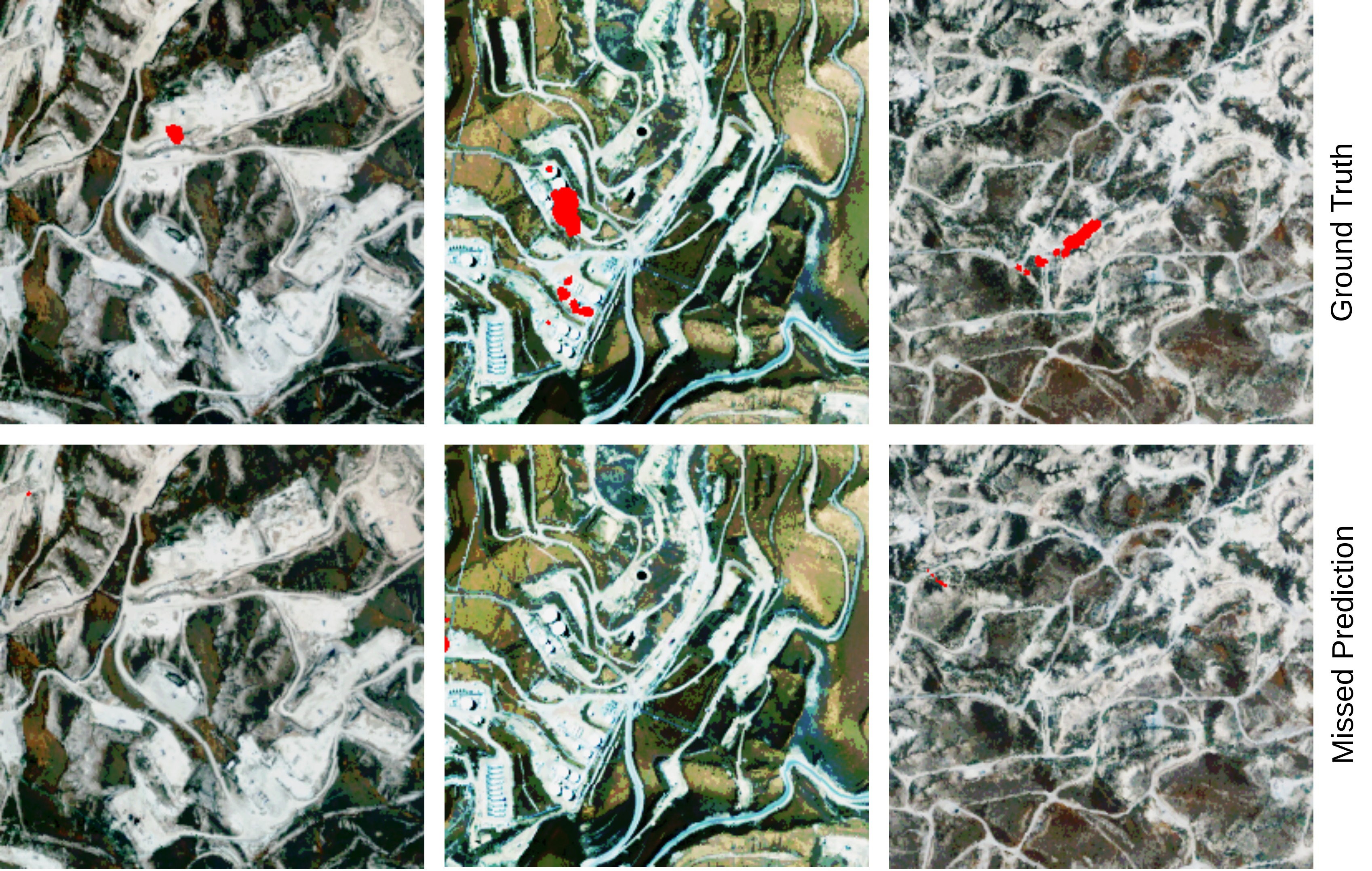}
\end{center}
\vspace{-0.45cm}
   \caption{\it Samples where MM fails to detect the CH$_4$ plume. We observed that these samples were recorded during the evening time and hence reflectance from the ground terrain is very weak. Therefore the absorption of reflected solar radiations by CH$_4$ is very low and hence the emissions goes undetected.}
\label{fig:mis_pred}
\end{figure*}

Figure~\ref{fig:mis_pred} shows some examples of missed CH$_4$ plume detections. We observed that going back to dataset samples and checking the timelines, these flightlines were recorded during the evening time. We believe that this might be because of evening time, the reflectance from the ground terrain is very weak and small. Hence we believe there is minimum absorption of reflected solar radiation by CH$_4$ gas present in the atmosphere and the plume goes undetected.

\subsection{Ablations Studies}
\noindent \textbf{Attention Type:} We also explored  different  attention mechanisms to encode and decode information. We replaced only the attention layers with deformable-attention~\cite{zhu2020deformable} in the our architecture that resulted in a drop of 0.1 mAP in the baseline model.

\subsubsection{Implementation details}
The whole network is trained with AdamW~\cite{loshchilov2017decoupled} optimizer, batch size of 12, with initial learning rate for backbones set to $10^{-5}$ and for transformer the learning rate is set to $10^{-5}$ with a weight decay of $10^{-4}$. The learning rate for mask prediction module is set to $10^{-4}$. The learning rate is dropped at every 150 epochs, we train for 300 epochs. The baseline model is trained on 2 V100 GPUs.

\clearpage

\end{document}